# Structural plasticity of single chromatin fibers revealed by torsional manipulation

Aurélien Bancaud[1], Natalia Conde e Silva[2], Maria Barbi[3], Gaudeline Wagner[1], Jean-François Allemand[4], Julien Mozziconacci[3], Christophe Lavelle[2,3], Vincent Croquette[4], Jean-Marc Victor[3], Ariel Prunell[2] and Jean-Louis Viovy[1]*

[1] Inst. Curie, UMR 168, Paris, 75231 France
[2] Institut Jacques Monod (UMR 7592), 2 place Jussieu 75251 Paris cedex05 France
[3] Laboratoire de Physique Théorique de la Matière Condensée (UMR 7600), 4 Place Jussieu 75252 Paris cedex05 France
[4] Laboratoire de Physique Statistique (UMR 8549), 24 Rue Lhomond 75231 Paris cedex05 France

* Corresponding author: jean-louis.viovy@curie.fr

**Magnetic tweezers are used to study the mechanical response under torsion of single nucleosome arrays reconstituted on tandem repeats of 5S positioning sequences. Regular arrays are extremely resilient and can reversibly accommodate a large amount of supercoiling without much change in length. This behavior is quantitatively described by a molecular model of the chromatin 3-D architecture. In this model, we assume the existence of a dynamic equilibrium between three conformations of the nucleosome, which are determined by the crossing status of the entry/exit DNAs (positive, null or negative). Torsional strain, in displacing that equilibrium, extensively reorganizes the fiber architecture. The model explains a number of long-standing topological questions regarding DNA in chromatin, and may provide the ground to better understand the dynamic binding of most chromatin-associated proteins.**

The genetic material of eukaryotic cells is organized into chromatin, a nucleoproteic structure whose repetitive unit is the nucleosome[1]. The core particle of the nucleosome consists of 147 bp of DNA wrapped 1.65 times around an octamer containing two copies each of the four core histones, H2A, H2B, H3 and H4[2]. This leads to both compaction and topological deformation of the DNA by one negative turn per nucleosome ($\Delta Lk \sim -1$, Lk being the linking number[3]). *In vivo*, regularly distributed nucleosome arrays with a repeat length of *ca.* 200 bp[1] fold into "30 nm fibers", whose modulated compaction is thought to be associated with a differential accessibility of DNA[4] to interactions with various factors, as required for DNA activity. A better knowledge of chromatin organization is expected, therefore, to improve our understanding of the regulation of DNA transactions *in vivo*.

*Bona fide* nucleosome arrays can be reconstituted *in vitro* and single molecule techniques now offer a direct approach to study their molecular dynamics in real time. Force micromanipulation has revealed the existence of an internucleosomal attraction that maintains the higher-order chromatin structure in physiological conditions[5] and a reversible peeling of ~80 bp of nucleosomal DNA below 15 pN[6], presumably accompanied by the destabilization of H2A-H2B dimers. Above this force, discrete disruption events of 25 nm each were observed, which were attributed to tetrasome ((H3-H4)$_2$-DNA complex) dissociation[6,7].

Here, we report the first investigation of the torsional response of single chromatin fibers using magnetic tweezers[8]. Nucleosome arrays reconstituted on 5S tandemly repeated positioning sequences were found to be able to accommodate large amounts of negative or positive supercoiling without much change in their length. A quantitative model is proposed, based on a dynamic equilibrium between the three conformations of the nucleosome previously identified through the minicircle approach (a single nucleosome reconstituted on a DNA minicircle)[9]. In these states, the nucleosome entry/exit DNAs can cross negatively (as in the canonical structure[2]), positively, or do not cross at all. The model fits the chromatin length-*vs.*-torsion response at different levels of compaction. It also shows how the torsional constraint, depending on its amplitude and sign, can force nucleosomes to switch conformation, and induce a large reorganization of the fiber architecture. These findings provide simple answers to long-standing topological mysteries of DNA in chromatin. Moreover, the dynamic chromatin it describes may underlie the dynamic nature of the binding of most chromatin-associated proteins[10,11].

**RESULTS**

**Torsion**

Nucleosome arrays were reconstituted by stepwise dilution using a linear DNA containing 36 tandemly repeated 208 bp 5S positioning sequences[12], and core histones purified from chicken erythrocytes. These fibers were then flanked by two naked DNA spacers, to avoid histone-mediated hydrophobic interaction with the surfaces, and by two "stickers" that link the fiber to the coated bottom of the flow cell and to the paramagnetic bead (blue and orange segments in **Fig. 1**). A pair of magnets was placed above this construction, and different torsions were applied by rotating the magnets about the vertical axis. The magnets' vertical position specifies the stretching force, i.e. the fiber extension, which was measured by recording the 3D-position of the bead[8].

The typical torsional behavior of a single chromatin fiber in low salt buffer B$_0$ (see Methods) is shown in **Fig. 2a** at 0.34 pN (blue curve). Following chemical dissociation of the nucleosomes, the response of the corresponding naked DNA was obtained (red). This latter curve displays a mechanical effect of torsion and an asymmetry for negative supercoiling, which are signatures of an unnicked single duplex DNA[8]. Compared to naked DNA, chromatin is shorter by ~1.35 μm, and its centre of rotation is shifted by −24±2 turns. This corresponds to a shortening of ~− 55 nm per negative turn, as expected for one nucleosome, since 50 nm correspond to 150 bp.

Nucleosomes were also disrupted mechanically by increasing the tension, after supplementing B$_0$ with 50 mM NaCl and 2×10$^{-3}$ % NAP-1 (Nucleosome Assembly

Protein-1, gift from S. Leuba). At 7.7 pN, 14 individual lengthening steps with an average height of 24.2±1.9 nm were detected (**Fig. 2b**), in agreement with[6]. This process occurred at a lower force than in[6], presumably because NAP-1 interacts with core histones *in vitro*[13] and favors their release. Interestingly, it was partially reversible, as also reported in[6]. In the course of two successive pulling phases at 7.7 pN, separated by a 50 sec pause at 0.67 pN, the fiber contracted by -28.5 nm during the pause, roughly corresponding to one individual reassociation.

The response in torsion of the partially disrupted fiber was subsequently probed in $B_0$ at low force and, as expected, found to be intermediate, in both rotation and length, between the responses of the original fiber and of DNA (**Fig. 2a**, green). The shifts in length and in topology between the new fiber and DNA were respectively ~700 nm and -13±1.5 turns, or ~54 nm per turn, identical to the above value. Assuming that each step corresponds to the dissociation of one nucleosome, then the topological deformation per nucleosome can be estimated to -(11±1.5)/14=-0.8±0.1 turn.

The rotational behavior of ten fibers was subsequently compared by plotting their maximal length at 0.3 ± 0.07 pN *versus* the rotational shift of those maxima relative to their corresponding naked DNA (**Fig. 2c**). A linear trend was observed, with most data points well-aligned and a rate close to 55 nm/turn. This is the expected behavior for regular nucleosome arrays with a variable number of nucleosomes. The corresponding nucleosome arrays were thus referred to as *regular*. A few fibers, however, deviated from this linear trend. We show in the Supplementary (**Fig. S1**) that these deviations can be attributed to the presence of variable proportions of clustered NSs devoid of linker DNA. Hence, these fibers were coined "*irregular*".

A direct comparison between chromatin and DNA requires to derive the torsional response of a DNA molecule of the same maximal length under the same force. Taking advantage of the invariance in length of the DNA torsional response[8], one can obtain the renormalized curve of the DNA by dividing both lengths and rotations by the ratio of the maximal length of DNA to the maximal length of chromatin. The DNA curve was further displaced parallel to the abscissa in order to superimpose the rotation centers (**Fig. 2d**). Compared to DNA, nucleosome arrays appear as extremely resilient, being able to accommodate a much larger amount of supercoiling than DNA without significant shortening.

## Stretching

The fiber described in **Fig. 2d** and its corresponding DNA were also compared, again in $B_0$, with respect to their stretching behaviors (**Fig. 3a**). Whereas chromatin is more rigid than DNA below 1 pN (the curve is steeper), i.e. in its entropic stretching regime (see below), it becomes more flexible beyond that force. This feature can be understood qualitatively as a consequence of the chromatin 3D-arrangement: bending a wire in a spring-like shape can reduce its stretching modulus by orders of magnitude, a property extensively used in engineering.

The dependence of the rotational behavior on the applied force was also studied (**Fig. 3b**; the force ranges from 0.09 pN to 0.34 pN). The torsional response always appeared more asymmetric at lower forces, and the curve's apex shifted towards negative values. This latter feature, which was not observed for naked DNA (not shown), indicates a larger topological deformation per nucleosome at lower force.

## Salt effects

Several buffer conditions were investigated, and two of them, representative of the general trend, are documented here: $B_0$ + 25 mM NaCl and $B_0$ + (40 mM NaCl + 2 mM MgCl$_2$). Compared to the $B_0$ results, the fibers systematically appeared more compact (by ~15% in the first conditions, and ~30 % in the second; **Figs. 4a** and **4b**, respectively). Interestingly, the condensed fiber in salt could be extended to the $B_0$ level by the transient application of a force of several pN. When the tension was released, the force-*vs.*-length behavior became virtually identical to that obtained in $B_0$. This property is visible in the force-vs.-length response: a hysteresis loop could be described as the fiber was always longer upon decreasing the force than upon increasing it (**Fig. 4c**). A similar behavior was observed upon rotation: if a force of ~2-3 pN was exerted immediately before a torsional manipulation (typically performed at 0.3 pN), the response of the fiber was nearly identical to that previously recorded in $B_0$ at the same force (data not shown).

The reference behavior in $B_0$ therefore seems to correspond to a maximal extension of the fiber, in agreement with earlier observations that nucleosome arrays are decondensed in low salt[14]. The condensation and the hysteretic behavior in higher salt conditions presumably reflect short-range attractive nucleosome-nucleosome interactions mediated by histone tails, which can be temporarily broken by a transient force increase. These results are quite consistent with those reported in[5], in which native chromatin was micromanipulated in tension under different salt conditions.

Despite a significant scattering intrinsic to those length measurements, a shift of the centre of rotation was always observed towards more negative values in higher salt (**Fig. 4b**), in a striking parallel with the behavior at lower force (see **Fig. 3b**). For instance, the shift from $B_0$ to $B_0$ + (40 mM NaCl + 2 mM MgCl$_2$) was -6 turns, i.e. ~-0.25 turn per nucleosome (**Fig. 4b**).

## Modeling

Worm-like rope approach and canonical chromatin The fiber's mechanical properties was first quantified using the worm-like rope model. This model, widely used for DNA[15], represents a molecule (or, here, the chromatin fiber) as an isotropic elastic rod with defined bending, twisting and stretching moduli. That model led to excellent fits of the length-*vs.*-torsion and force-*vs.*-length responses in **Figs. 2d** and **3a**. The fitted values of the bending persistence length and stretching modulus (28 nm and 8 pN, respectively) were in agreement with previous studies[5,6] and models of the chromatin fiber[16,17]. In contrast, the torsional persistence length, ~5 nm (against ~80 nm for DNA), was exceptionally low.

As a first attempt to interpret this torsional resilience, we modeled two-angle nucleosomes[18,19] with their entry-exit DNAs crossed negatively, as inferred from the core particle crystal structure[2] and observed with crystallized tetranucleosomes[20]. Connecting these canonical nucleosomes by flexible DNA linkers led to the "all-negative" fiber in Supplementary (**Fig. S2**). Its very large torsional persistence length (35 nm, against 5 nm for the experimental value; see above) prompted us to turn to other concepts.

<u>Mononucleosomes assembled on DNA minicircles</u> Previous studies demonstrated that mononucleosomes thermally fluctuate between three discrete conformational states corresponding to different entry-exit DNAs crossing status (with $\Delta Lk_o \sim$ -0.7 (open: no crossing), $\Delta Lk_n \sim$ -1.4 (closed negative: negative crossing) and $\Delta Lk_p \sim$ -0.4 (closed positive: positive crossing) for the 5S positioning sequence (**Fig. 5a**)). The transition between these states involves a rotation of the nucleosome around its dyad relative to the loop[9,21]. Notably, cryo-electron microscopic visualization of reconstituted fibers in low salt conditions also suggested the occurrence of such states[22]. This differs from the tetranucleosome crystal structure[20] presumably because the high salt conditions used for crystallization favor closed conformations[9].

The existence of the open state was first documented in minicircles[23], but it was only after the core particle crystal structure was disclosed[2], that the reason for such an easy unwrapping of the nucleosome edges became clear. DNA is attached to the octamer at 14 specific binding sites. These 14 sites are spaced every ~10 bp, and are defined by their Super Helix Location (SHL) relative to the dyad[2]. The SHL±6.5 sites are located at the nucleosome entry/exit, and have the weakest binding energy[24]. The existence of discrete open and closed states therefore results from the status of these sites, which can only be *on* or *off*.

The closed positive state shows a positive crossing of entry-exit DNAs. Although counter-intuitive given the left-handed wrapping around the histone octamer, this state has been extensively documented using the minicircle approach through ethidium bromide fluorescence titration[25] and relaxation[21,26,27]. We recently confirmed that SHL±6.5 binding sites are *on* in this closed positive conformation. Indeed, the substitution of H3 arginine 49 by a lysine, which in contrast to arginine cannot intercalate its basic lateral chain into the small groove of the DNA[2], equally affects the energies of the closed negative and closed positive states (N. C. e S. & A. P., unpublished results).

<u>The model</u>. Our molecular model for the fiber (208 bp repeat length) similarly assumed a thermodynamic equilibrium between the three different states of the nucleosome (**Fig. 5a**). For comparison, "all-open" and "all-positive" fibers are shown in the Supplementary material (**Fig. S3**). A standard statistical mechanical analysis (free energy minimization based on the partition function) could then predict the fiber length-*vs.*-torsion behavior at constant force, as a function of the energy differences between the states (see details in Supplementary Material §2). The upper part of the rotational response of a *regular* nucleosome array corresponding to the blue arrowhead in **Fig. 2c** was accurately fitted by this model (**Fig. 5b**, bold line), using the number of nucleosomes (31), and the energy differences between the negative (respectively positive) and the open state (+0.7 kT, respectively +2 kT) as adjustable parameters. The low energies involved insure that nucleosomes in the fiber are in a dynamic equilibrium, and this equilibrium is displaced by the applied torsion. Fiber 1 in **Fig. 5b** has the maximal extension and most of its nucleosomes in the open state, whereas fibers 2 and 3 are slightly shorter and have most, if not all of their nucleosomes in the negative and positive states, respectively. The model also provides a prediction of the torque as a function of torsion (**Fig. 5c**) (a similar curve was actually obtained when considering the fiber as an isotropic elastic rod;

Supplementary **Fig. S4**). The torque is less than 3 pN•nm/rad over ~30 turns around the center of rotation, i.e. significantly smaller than that exerted by polymerases (>5 pN•nm/rad[28]) or than the value predicted for nucleosome torsional ejection (9 pN•nm/rad)[29].

Remarkably, the behavior of all *regular* nucleosome arrays could be described with the same set of energy values by fitting the number of nucleosomes only (**Fig. 6**). The fitted energies (0.7 and 2 kT; see above) are close to those obtained in the minicircle system under conditions of maximal entry-exit DNAs repulsion (0.8 and 3.6 kT[9]). Considering the differences in geometry and ionic environment between the two systems, our best fit values appear to be fully consistent with minicircle data.

Response to high torsional stress Once ~-20 turns have been applied starting from the apex, the model predicts that all nucleosomes should be in the closed negative state (fiber 2 in **Fig. 5b**). Because the fiber in the model cannot accommodate more negative turns, the torque then increases abruptly (dotted line in **Fig. 5c**). In practice, a marked change in the length-*vs.*-rotation curve is observed beyond ~-20 turns, which reflects the transition to a regime with a constant slope of ~-25 nm/turn. By analogy with DNA[8], we interpret this constant slope as a consequence of plectoneme formation. Interestingly, 25 nm/turn are significantly smaller than the 90 nm/turn obtained for DNA at the same force, which indicates a lower torque (3 pN.nm/rad for chromatin, against 6 pN.nm/rad for DNA, see Supplementary Material §3). Hence, plectonemes may not form in the DNA spacers flanking the nucleosome array (**Fig. 1**), but rather in the fiber itself, this process presumably extruding the nucleosomes away from the plectoneme axis (Supplementary **Fig. S5**). It is noteworthy that plectoneme formation may be facilitated in chromatin for two reasons. The energy cost of bending DNA, which is critical for naked DNA, is reduced by nucleosomes, which are natural DNA benders (also note that the persistence length of the fiber is smaller than that of DNA, see above). Second, the DNA charge screening of entry/exit DNAs through interaction of H3 N-terminal tails[30,31] may reduce their effective diameter.

Plectonemes should also develop on the positive side of the rotation curve, and indeed a transition to a linear slope of ~25 nm/turn occurs at ~ +10 turns from the apex, i.e. at a significantly lower supercoiling than on the negative side. Our model predicts that the corresponding torque should still be ~3 pN.nm/rad (see Supplementary Material §3), a value similar to that obtained on the negative side. Moreover, only a fraction of the nucleosomes are expected to be in the closed positive state in the corresponding fiber (fiber 3 in **Fig. 5b**). Consistent with an energetically unfavorable closed positive state, it follows that the torque necessary to drive the fiber to an "all-positive" state should be higher than the critical torque for plectonemes formation.

## DISCUSSION

### Nucleosome transitions and chromatin topology

These nanomanipulation experiments show that regular chromatin fibers are torsionally resilient structures that can accommodate large positive and negative supercoiling without developing strong torque or undergoing significant shortening.

This resilience, typically 5 times higher than predicted for a canonical fiber of closed negative nucleosomes, was interpreted as being the consequence of a dynamic equilibrium occurring between three conformational states of the nucleosomes. A molecular model based on this equilibrium quantitatively accounts for the data, and unravels the energy landscape involved in those nucleosome transitions.

This dynamic nature of chromatin provides simple explanations to several long-standing puzzles about the topology of DNA in chromatin. The most well-known is the so-called "linking number paradox": why does a two-turn particle reduce the DNA linking number by one, instead of two[9,32,33]? At first, the DNA was proposed to become overtwisted upon wrapping on the histone surface[32], but it was later recognized that, if some overtwisting may indeed occur, it was by no way sufficient to explain this discrepancy[4]. The true explanation may, therefore, lie in the dynamic topological compensation occurring between negatively and positively crossed nucleosomes.

Several other pending questions find their simple answers. i) The shift of the $\Delta Lk$ per nucleosome observed for a minichromosome reconstituted on the same 5S repeats from -1.0 with control histones to -0.8 with hyperacetylated histones (i. e. under high mutual repulsion of linker DNAs)[34] was clearly due to a displacement of the dynamic equilibrium toward more nucleosomes in the open state. The same occurred in our experiment (**Fig. 4b**): $\Delta Lk \sim$ -1 obtained in higher salt against $\Delta Lk \sim$ -0.8 in $B_0$ (also favoring entry-exit DNA repulsion) [Interestingly, a similar $\sim$+0.2 shift in $\Delta Lk$ was also measured in the minicircle system with acetylated mononucleosomes in phosphate, a buffer which further destabilizes histone tails/DNA interactions[9].] ii) The property of reconstituted minichromosomes to withstand as much negative supercoiling ($\sigma \sim$-0.1) as the corresponding naked DNA upon treatment with DNA gyrase[35], i.e. the nucleosome apparent "transparency" to that enzyme in spite of the trapping of most DNA in the nucleosome cores, must result from the shift of all nucleosomes to the negative state[9], rather than from a forced undertwisting of the DNA on the histone surface[34]. iii) Finally, the ability of positively supercoiled plasmids to reconstitute a large number of nucleosomes, without apparent interference of the large additional positive supercoiling which was expected to accumulate[36], must similarly reflect nucleosomes displacement toward the positive state.

**Chromatin as a topological buffer**

One may question the biological relevance of conclusions about the topology of nucleosomes drawn from experiments performed on chromatin fibers devoid of linker histones. Linker histones cannot bind nucleosomes in the open state, but they do bind nucleosomes in the negative and positive states, and this brings entry/exit DNAs together into a torsionally highly flexible stem[27]. Thus, even if the steady-state proportion of open-state nucleosomes is small *in vivo*, at least in quiescent chromatin, the H1-containing fiber with nucleosomes in the closed negative and closed positive states should remain highly resilient. Open nucleosomes may rather be more involved in active chromatin, as suggested by two observations. First, histone acetylation, which favors the open state (see above), is usually associated with transcription. Second, H2A-H2B dimers were much more readily removed by NAP-1 when nucleosomes were in the open state than in the negative state, that removal resulting in a further unwrapping and the formation of single-turn tetrasomes (N. C. e S. & A. P., unpublished results).

Chromatin torsional resilience must have important *in vivo* implications because DNA transactions usually involve topological changes. Chromatin is first expected to act as an efficient damper against torsional waves generated by tracking enzymes[37], which may favor their smooth progression by avoiding the formation of plectonemic structures that may affect the chromatin large scale spatial arrangement, and protect nucleosomes from unsolicited destruction by positive supercoiling. For instance, the three-states model described here predicts that a chromatin fiber with 40 nucleosomes and fixed ends is able to withstand the supercoiling generated by the transcription of circa 100 bp without the help of topoisomerases, and without exceeding the torque exerted by the polymerase (see **Fig. 7** and its legend for details).

Nucleosome conformational transitions may also play a role in the control of DNA-protein interactions in the chromatin context, for instance by affecting the binding of linker histones and their HMG proteins competitors[38], and probably also of other proteins such as remodeling or transcription factors[11] by means of rearrangements in the fiber 3D architecture (see e.g., **Fig. 5b and 7**).

As suggested in[38], dynamic binding of proteins on chromatin[10,11,39] offers an efficient way to quickly react to changes in the environment. Because they depend on the fiber's torsion, the transitions revealed and discussed in this work may provide the conditions for a coupling between this dynamic binding and the action of tracking enzymes. This coupling has the rather unique property of being both long-range and much faster than any molecular transport process. It is thus a particularly interesting candidate for fast-responding regulatory mechanisms.

## METHODS

**Nucleosome arrays preparation**. Nucleosome arrays were reconstituted by conventional stepwise dilution. The nucleosome density was checked by sedimentation in sucrose gradients[40] and the nucleosome array regularity probed by microccocal nuclease digestion (not shown).

Three DNA fragments were prepared by PCR. Two of them were amplified from the linearized template Litmus28i (NEB, position 2008 and 2580) with modified biotin or digoxigenin nucleotides (Roche). The third one was obtained by amplifying the pFOS-1 template (NEB, position 3803 and 4539) with standard nucleotides. Appropriate restriction digestions of the PCR products led to 554 and 620 bp fragments. These fragments were ligated into two different 1174 bp "hybrids", consisting of one part (620 bp) of unmodified DNA and another part (574 bp) modified with biotin or digoxigenin.

The two "hybrid" fragments were then ligated to the nucleosome arrays to give the final construction (**Fig. 1**). The fibers were finally dialysed against TE (10 mM Tris-HCl, pH=7.5 and 1 mM EDTA), and stored at −20°C following a two-fold dilution with 100% glycerol.

**Magnetic tweezers apparatus**. A poly-dimethylsiloxane (PDMS) (Dow-Corning) flow cell with a 2 mm wide and 80 μm high channel was constructed. This microfluidic cell was mounted on a glass coverslip treated with 3-mercaptopropyl-trimethoxysilane (Sigma)[41]. The surface coating was performed inside the channel with non-specific binding of anti-digoxigenin (Roche) during 1 hour at 37°C, followed by overnight BSA blocking.

The PDMS flow cell was placed beneath two NdFeB permanent magnets (HPMG) separated by 0.8 mm[41]. Images were grabbed by a CCD camera (JAI). From the transverse fluctuations magnitude and the molecule length, the exact force acting on the bead was deduced[8]. Moving the magnets up and down by ~5 mm permits a range of forces from 0.1 pN to 15 pN. The topological constraint was controlled by rotation of the magnets about the vertical axis.

**Nucleosome array injection and study**. Just prior to the experiment, 1 ng of chromatin, previously diluted to 10 μL with TE, was mixed with 100 μg of 2.8 μm diameter streptavidin-coated magnetic beads (Dynal). After 1 minute of incubation, the solution was aspirated into the cell by a syringe pump. Data were usually acquired in TE plus 0.01 % BSA (Bovine Serum Albumin) ($B_0$). The standard buffer was $B_0$ because, at this low ionic strength, nucleosomes are very stable and do not move along DNA[42], and nucleosome/nucleosome interactions are weak[43].

**Chemical nucleosome disruption**. At the end of each experiment, nucleosomes were chemically disassembled by aspirating into the flow-cell a solution containing 5% heparin (Dakota Pharm) in $B_0$ during 10 minutes.

ACKNOWLEDGMENTS
The authors are grateful to G. Almouzni and J.P. Quivy for providing material in preliminary experiments, E. Ben-Haim and C. Bouchiat for discussions, K. Dorfman and A. Sivolob for a critical reading. AB, NC-S and JM thank the French ministry of research for AC-MENRT fellowships. This work was supported by CNRS (AP and JMV labs), and by grants from the CNRS/MENRT programs "DRAB" and Institut Curie cooperative program "Physics of the cell" (JLV lab).

Author information:
Reprints and permissions information is available at npg.nature.com/reprintsandpermissions. The authors declare no competing financial interests. Correspondence and requests for materials should be addressed to JLV (jean-louis.viovy@curie.fr).

**Figure 1** Schematics of the experiment. A single nucleosome array (~7.5 kbp), sandwiched between two naked DNA spacers (~600 bp each), is linked to a coated surface and to a magnetic bead. A pair of magnets placed above this molecule exerts controlled torsional and extensional constraints[8].

**Figure 2** Micromanipulation of single chromatin fibers and DNA. **(a)** Extension-*vs.*-rotation curves at 0.35 pN for an intact fiber (blue) in buffer $B_0$ (see Methods), for the same fiber after partial nucleosome disruption in **b** (green), and for its naked DNA after complete nucleosome dissociation (red). **(b)** Individual nucleosome disruption events at 7.7 pN of the fiber in **a** in $B_0$ plus NAP-1 and 50 mM salt (see Results). The force is temporarily lowered to 0.67 pN between the arrows. **(c)** Maximal extension-vs.-topological departure from DNA for 10 fibers at $0.3 \pm 0.07$ pN in $B_0$. The black straight line is the relationship predicted by our 3-state model (see Supplementary). Fibers on that straight line are referred to as *regular*, and those off as *irregular*. Arrowheads correspond to the fiber studied in **a** (black) and **d** (blue). Numbers in green refer to the fibers studied in **Fig. 6**, below. **(d)** Extension-*vs.*-rotation curve of the chromatin fiber (blue) corresponding to the blue arrowhead in **c** at 0.25 pN with its corresponding renormalized DNA (red, see text). Smooth curves for the fiber and naked DNA were obtained assuming an elastic response in bending, stretching and twisting (worm-like rope model)[15].

**Figure 3** Tension-dependence of the fiber mechanical behavior. **(a)** Force-*vs.*-extension curves in $B_0$ of the fiber in **Fig. 2c** and **2d** (blue) and of its DNA (red) at their respective centers of rotation. Smooth curves were obtained as described in legend to **Fig. 2d**. **(b)** Extension-vs.-torsion for the fiber in **Fig. 2a** (blue) in $B_0$ under tensions of 0.09 pN (triangles), 0.17 pN (circles), and 0.34 pN (crosses). A strong asymmetry in the mechanical response for positive *vs.* negative torsional constraints is observed at low forces, which we attribute to the different energies of the nucleosome positive and negative states. The apex also shifts towards negative torsion at lower forces, presumably as a consequence of a shift in the equilibrium toward more nucleosomes in the negative conformation.

**Figure 4** Salt-dependence of the fiber mechanical behavior. **(a)** Extension-*vs.*-rotation behavior of a fiber in $B_0$ (black) and in $B_0$ + 25 mM NaCl (blue). **(b)** Extension-*vs.*-rotation behavior of a fiber in $B_0$ (solid black line) and in $B_0$ + (40 mM NaCl + 2 mM MgCl$_2$) (blue crosses, filtered average: blue line). An increased variability in the measurements is observed and the apex of the average curve shifts towards more negative rotation values, reflecting a displaced equilibrium with more nucleosomes in the negative state (see text). **(c)** Response of the fiber in **a** in force-*vs.*-extension at its centre of rotation. In this experiment, we describe "force-cycles": first, the distance in between the magnets and the fiber is progressively lowered, and for each step (i.e. each force) the length of the fiber is recorded (dark colors). Once a constraint of ~5 pN is reached, the process is reversed, and the force is progressively lowered down to its initial value (~0.05 pN, light colors). The fiber does not show any hysteresis in $B_0$ (black and grey), in contrast to the fiber in $B_0$ + 25 mM NaCl (dark and light blue), which is initially more compact, but can be extended to the $B_0$ level by a force of a few pN.

**Figure 5** The model. **(a)** Representation of individual nucleosomes in the negative ($\alpha \sim 54°$, blue), open ($\alpha \sim -30°$, yellow) and positive ($\alpha \sim 30°$, cyan) states.

**(b)** Torsional data from **Fig. 2d** (squares) fitted by our three states nucleosome model (bold line). The model fits the response over 30 turns around the apex (bold smooth curve). For higher torsion (on the positive and negative sides) a thin line representing the best-fit plectoneme model is plotted. Typical structures of a 208 bp repeat fiber at the apex (circle 1; 65%, 20% and 15% of the nucleosomes are on average in the open, positive and negative conformations, respectively), and at the transition to the plectoneme regime on the negative side (circle 2; 100% negative nucleosomes, on average) or on the positive side (circle 3) (80% positive and 20% open nucleosomes on average) are drawn below the curve. **(c)** Torque as predicted by the three-states model (bold line) and by the plectoneme model (thin lines). The circles figure the transition regions between the two more or less overlapping regimes.

**Figure 6** Single parameter fitting. The molecular model assumes that $U_n$ and $U_p$ (0.7 and 2 kT, respectively) do not change from fiber to fiber, and a single-parameter fitting (adjusting the number of nucleosomes) can subsequently be performed. Here, the torsional behaviors of four *regular* fibers are plotted (blue, green, purple and red data points at $0.3 \pm 0.07$ pN in $B_0$), together with their best-parameter fits (black smooth curves), corresponding to the number of nucleosomes indicated. These responses correspond to the fibers 1-4 in **Fig. 2c**.

**Figure 7** Chromatin as a topological buffer. Schematics of the twin-supercoiled domain model of transcription[37] adapted to chromatin. Assuming that the fiber has clamped ends and that the transcription machinery cannot rotate around the helical axis of chromatin, the progression of the enzyme inside generates a positive torsional stress ahead of it (on the left side), and a negative one in its wake (on the right side). *In vivo*, immobilization of the fiber ends can be insured by proteins, as in chromatin loops, or by the viscous drag[44]. The blue box figures the transcription bubble without any intended assumptions on the fate of nucleosomes under transcription. (**a**) At the onset of transcription, the whole fiber is torsionally relaxed (Supplementary **Fig. S4**). Once started, the polymerase will keep moving until the mounting torques exerted by the left and right parts of the fiber balance the torque generated by the enzyme. For example, the transcription of 100 bp induces ~10 positive turns on the right part of the fiber, and ~10 negative on the left side. The Lk difference between a relaxed and an "all-negative" or a "most-positive" fiber is ~-0.6 or ~+0.4 per nucleosome, respectively (see **Fig. 5b** and Supplementary Material). For these two constrained states of the fiber, the torque is ~3 pN.nm/rad. Hence, the total torque is ~6 pN.nm/rad, close to the torque exerted by the polymerase, at least 5 pN.nm/rad[28]. We conclude that a fiber containing ~40 nucleosomes (equivalently, a topological buffer of ~40*0.5=20 turns) can sustain the transcription of 100 bp with no need for relaxation by a topoisomerase. Notably, because this topological buffer property involves transitions at the nucleosome level (a ~6 nm particle), we expect the viscous drag effects to be considerably

smaller than on the whole "30-nm fiber". The chromatin 3D-reorganization could therefore be propagated over longer fragments. (**b**) At the end of the elongation phase, the left part of the fiber is in a "most-positive" state (fiber 3 in **Fig. 5b**), whereas the right part is in an "all-negative" state (fiber 2 in **Fig. 5b**).

1.  Van Holde, K.E. *Chromatin*, (Springer-Verlag, New York, 1988).
2.  Luger, K., Mader, A.W., Richmond, R.K., Sargent, D.F. & Richmond, T.J. Crystal structure of the nucleosome core particle at 2.8 A resolution. *Nature* **389**, 251-260 (1997).
3.  Germond, J.E., Hirt, B., Oudet, P., Gross-Bellark, M. & Chambon, P. Folding of the DNA double helix in chromatin-like structures from simian virus 40. *Proc. Natl. Acad. Sci. USA* **72**, 1843-1847 (1975).
4.  Wolffe, A. *Chromatin*, (Academic Press, London, 1998).
5.  Cui, Y. & Bustamante, C. Pulling a single chromatin fiber reveals the forces that maintain its higher-order structure. *Proc. Natl. Acad. Sci. USA* **97**, 127-132. (2000).
6.  Brower-Toland, B.D. et al. Mechanical disruption of individual nucleosomes reveals a reversible multistage release of DNA. *Proc. Natl. Acad. Sci. USA* **99**, 1960-1965 (2002).
7.  Hayes, J.J. & Hansen, J.C. New insights into unwrapping DNA from the nucleosome from a single-molecule optical tweezers method. *Proc. Natl. Acad. Sci. USA* **99**, 1752-1754 (2002).
8.  Strick, T.R., Allemand, J.F., Bensimon, D., Bensimon, A. & Croquette, V. The elasticity of a single supercoiled DNA molecule. *Science* **271**, 1835-1837 (1996).
9.  Prunell, A. & Sivolob, A. Paradox lost : nucleosome structure and dynamics by the DNA minicircle approach. in *Chromatin Structure and Dynamics: State-of-the-Art*, Vol. 39 (eds. Zlatanova, J. & Leuba, S.H.) 45-73 (Elsevier, London, 2004).
10. Catez, F. et al. Network of dynamic interactions between histone H1 and high-mobility-group proteins in chromatin. *Mol Cell Biol* **24**, 4321-8 (2004).
11. Phair, R.D. et al. Global nature of dynamic protein-chromatin interactions in vivo: three-dimensional genome scanning and dynamic interaction networks of chromatin proteins. *Mol Cell Biol* **24**, 6393-402 (2004).
12. Simpson, R.T., Thoma, F. & Brubaker, J.M. Chromatin reconstituted from tandemly repeated cloned DNA fragments and core histones: a model system for study of higher order structure. *Cell* **42**, 799-808 (1985).
13. McBryant, S.J. et al. Preferential binding of the histone (H3-H4)2 tetramer by NAP1 is mediated by the amino-terminal histone tails. *J. Biol. Chem.* **278**, 44574-44583. (2003).
14. Leuba, S.H. et al. Three-dimensional structure of extended chromatin fibers as revealed by tapping-mode scanning force microscopy. *Proc Natl Acad Sci U S A* **91**, 11621-11625 (1994).
15. Bouchiat, C. & Mezard, M. Elasticity model of a supercoiled DNA molecule. *Phys. Rev. Lett.* **80**, 1556-1559 (1998).

16.  Ben-Haim, E., Lesne, A. & Victor, J.M. Chromatin: a tunable spring at work inside chromosomes. *Phys. Rev. E Stat. Nonlin. Soft Matter Phys.* **64**, 051921 (2001).

17.  Schiessel, H., Gelbart, W.M. & Bruinsma, R. DNA folding: structural and mechanical properties of the two-angle model for chromatin. *Biophys J* **80**, 1940-1956 (2001).

18.  Woodcock, C.L., Grigoriev, S.A., Horowitz, R.A. & Whitaker, N. A chromatin folding model that incorporates linker variability generates fibers resembling native structures. *Proc. Nat. Acad. Sci. USA* **90**, 9021-9025 (1993).

19.  Barbi, M., Mozziconacci, J. & Victor, J.M. How the chromatin fiber deals with topological constraints? *Phys. Rev. E Stat. Nonlin. Soft Matter Phys.* **71**, 031910 (2005).

20.  Schalch, T., Duda, S., Sargent, D.F. & Richmond, T.J. X-ray structure of a tetranucleosome and its implications for the chromatin fiber. *Nature* **436**, 138-141 (2005).

21.  De Lucia, F., Alilat, M., Sivolob, A. & Prunell, A. Nucleosome dynamics. III. Histone tail-dependent fluctuation of nucleosomes between open and closed DNA conformations. Implications for chromatin dynamics and the linking number paradox. A relaxation study of mononucleosomes on DNA minicircles. *J. Mol. Biol.* **285**, 1101-1119 (1999).

22.  Bednar, J. et al. Nucleosomes, linker DNA, and linker histone form a unique structural motif that directs the higher order folding and compaction of chromatin. *Proc. Natl. Acad. Sci. USA* **95**, 14173-14178 (1998).

23.  Goulet, I., Zivanovic, Y., Prunell, A. & Revet, B. Chromatin reconstition on small DNA rings. I. *J Mol Biol* **200**, 253-66 (1988).

24.  Luger, K. & Richmond, T.J. DNA binding within the nucleosome core. *Curr Opin Struct Biol* **8**, 33-40. (1998).

25.  Sivolob, A., De Lucia, F., Revet, B. & Prunell, A. Nucleosome dynamics. II. High flexibility of nucleosome entering and exiting DNAs to positive crossing. An ethidium bromide fluorescence study of mononucleosomes on DNA minicircles. *J Mol Biol* **285**, 1081-99. (1999).

26.  Sivolob, A., Lavelle, C. & Prunell, A. Sequence-dependent nucleosome structural and dynamic polymorphism. Potential involvement of histone H2B N-terminal tail proximal domain. *J Mol Biol* **326**, 49-63. (2003).

27.  Sivolob, A. & Prunell, A. Linker histone-dependent organization and dynamics of nucleome entry/exit DNAs. *J. Mol. Biol.* **331**, 1025-1040 (2003).

28.  Harada, Y. et al. Direct observation of DNA rotation during transcription by Escherichia coli RNA polymerase. *Nature* **409**, 113-115 (2001).

29.  Sarkar, A. & Marko, J.F. Removal of DNA-bound proteins by DNA twisting. *Phys Rev E Stat Nonlin Soft Matter Phys* **64**, 061909 (2001).

30.  Shaw, S.Y. & Wang, J.C. Knotting of a DNA chain during ring closure. *Science* **260**, 533-536 (1993).

31.  Angelov, D., Vitolo, J.M., Mutskov, V., Dimitrov, S. & Hayes, J.J. Preferential intreraction of the core histone tail domains with linker DNA. *Proc Natl Acad Sci USA* **98**, 6599-6604 (2001).

32.  Klug, A. & Lutter, L.C. The helical periodicity of DNA on the nucleosome. *Nucleic Acids Res* **9**, 4267-83 (1981).

33. Prunell, A. A topological approach to nucleosome structure and dynamics: the linking number paradox and other issues. *Biophys J* **74**, 2531-2544 (1998).

34. Norton, V.G., Imai, B.S., Yau, P. & Bradbury, E.M. Histone acetylation reduces nucleosome core particle linking number change. *Cell* **57**, 449-57 (1989).

35. Garner, M.M., Felsenfeld, G., O'Dea, M.H. & Gellert, M. Effects of DNA supercoiling on the topological properties of nucleosomes. *Proc. Natl. Acad. Sci. USA* **84**, 2620-2623 (1987).

36. Clark, D.J., Ghirlando, R., Felsenfeld, G. & Eisenberg, H. Effect of positive supercoiling on DNA compaction by nucleosome cores. *J. Mol. Biol.* **234**, 297-301. (1993).

37. Liu, L.F. & Wang, J.C. Supercoiling of the DNA template during transcription. *Proc. Natl. Acad. Sci. USA* **84**, 7024-7027 (1987).

38. Bustin, M., Catez, F. & Lim, J.H. The dynamics of histone H1 function in chromatin. *Mol Cell* **17**, 617-620 (2005).

39. Misteli, T., Gunjan, A., Hock, R., Bustin, M. & Brown, D.T. Dynamic binding of histone H1 to chromatin in live cells. *Nature* **408**, 877-881 (2000).

40. Hansen, J.C. & Lohr, D. Assembly and structural properties of subsaturated chromatin arrays. *J. Biol. Chem.* **268**, 5840-5848. (1993).

41. Fulconis, R. et al. Twisting and untwisting a single DNA molecule covered by RecA protein. *Biophys J* **87**, 2552-2563. (2004).

42. Meersseman, G., Pennings, S. & Bradbury, E.M. Mobile nucleosomes--a general behavior. *EMBO J.* **11**, 2951-2959. (1992).

43. Mangenot, S., Leforestier, A., Durand, D. & Livolant, F. X-ray diffraction characterization of the dense phases formed by nucleosome core particles. *Biophys J* **84**, 2570-2584 (2003).

44. Mondal, N. et al. Elongation by RNA polymerase II on chromatin templates requires topoisomerase activity. *Nucleic Acids Res* **31**, 5016-5024 (2003).

# Figure 1

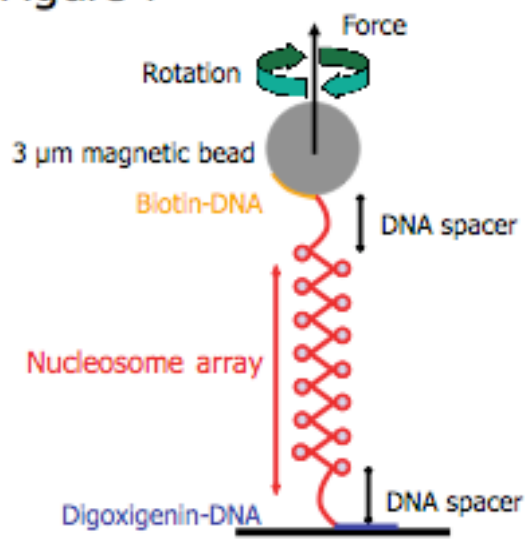

# Figure 2

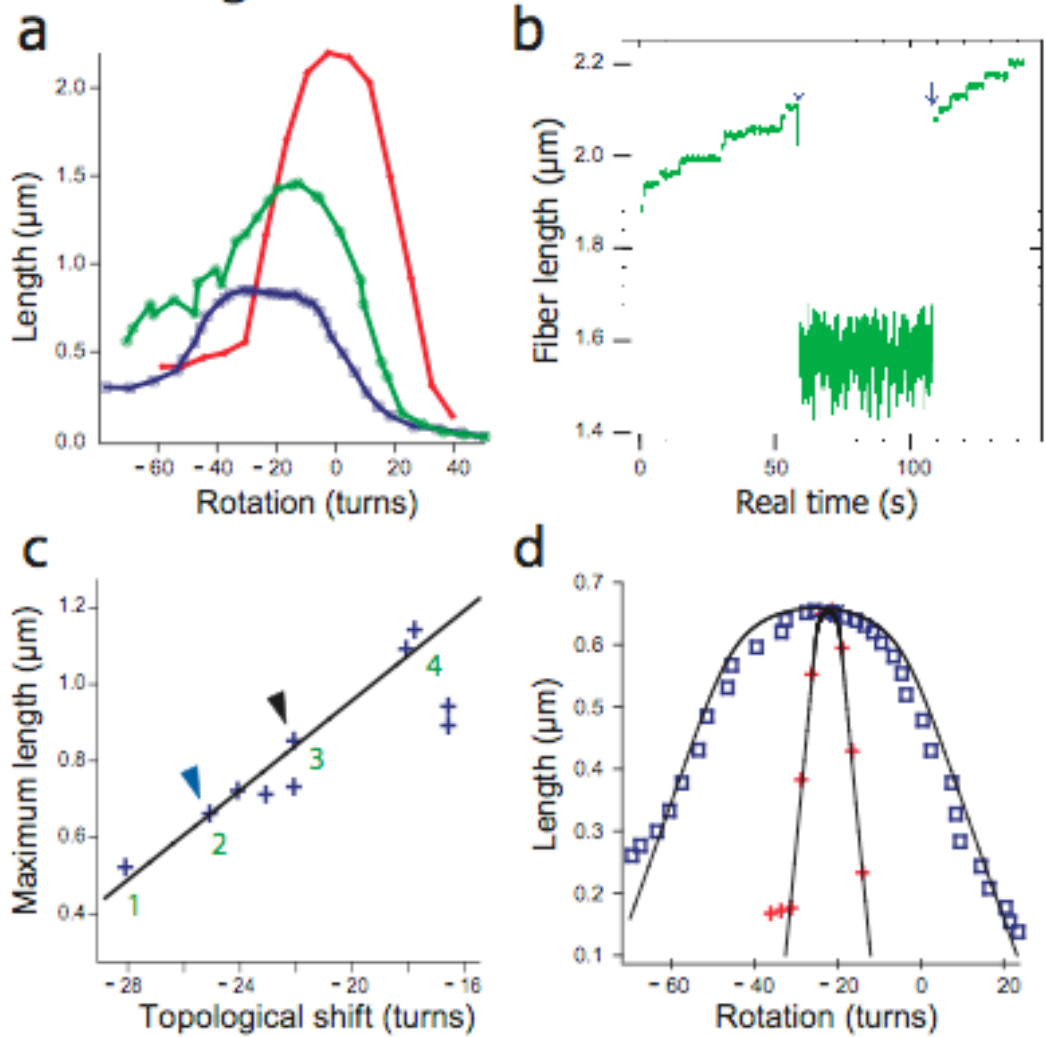

## Figure 3

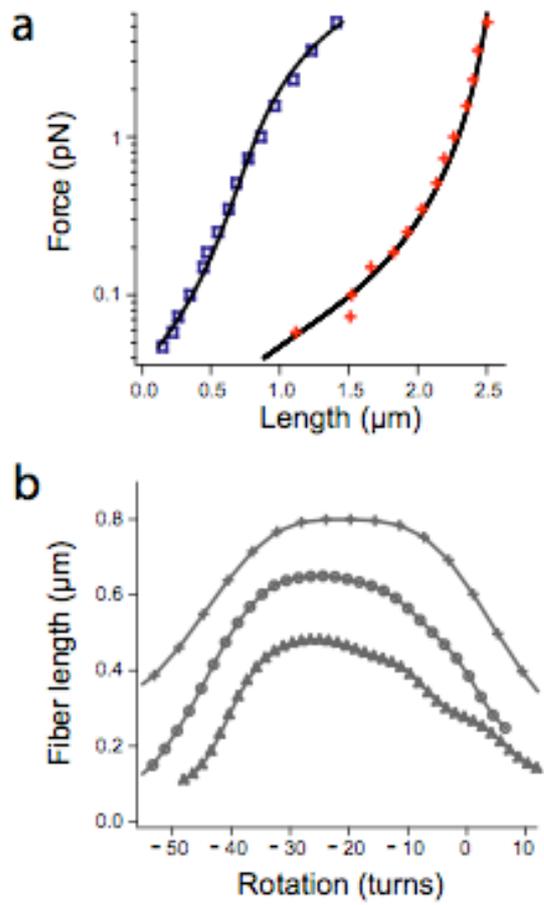

## Figure 4

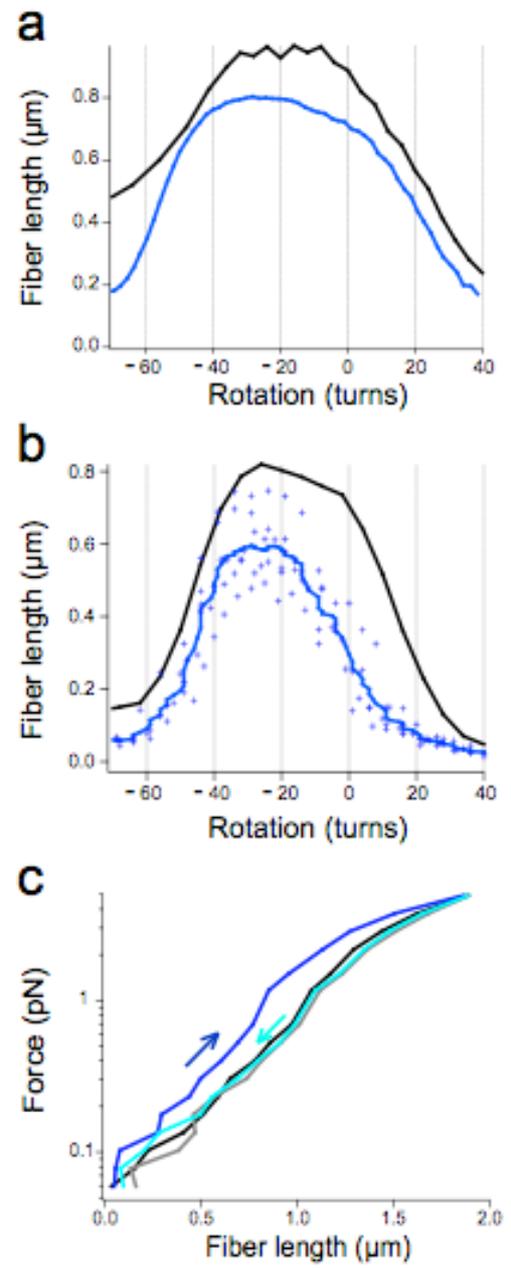

# Figure 5

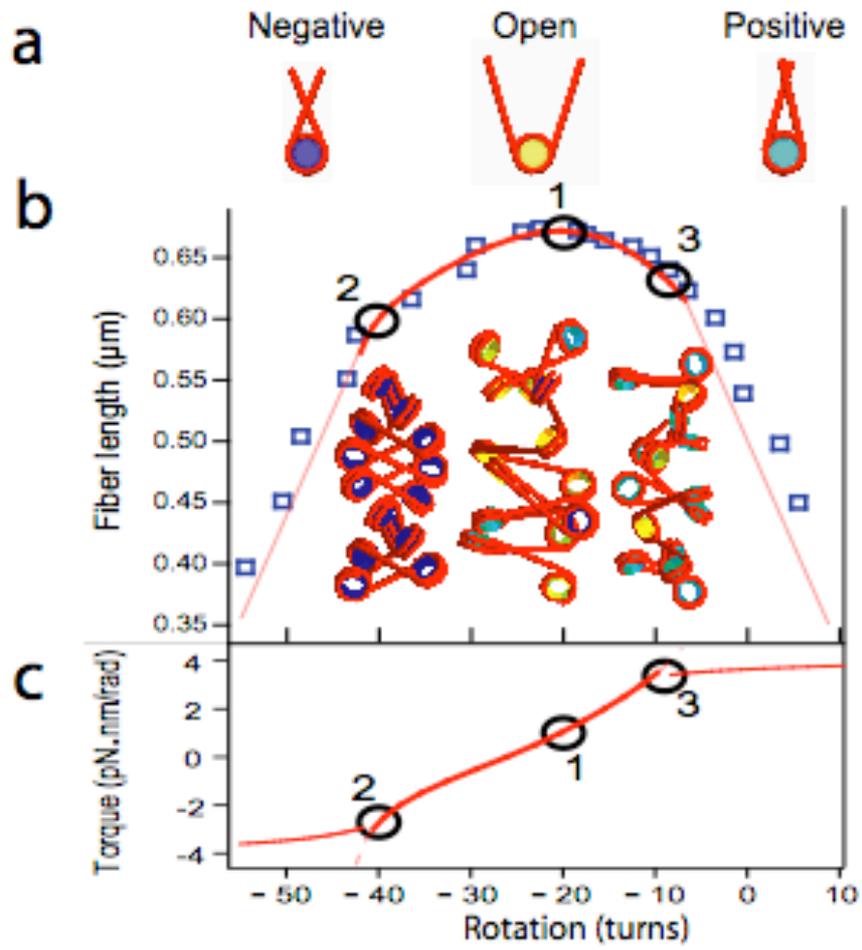

# Figure 6

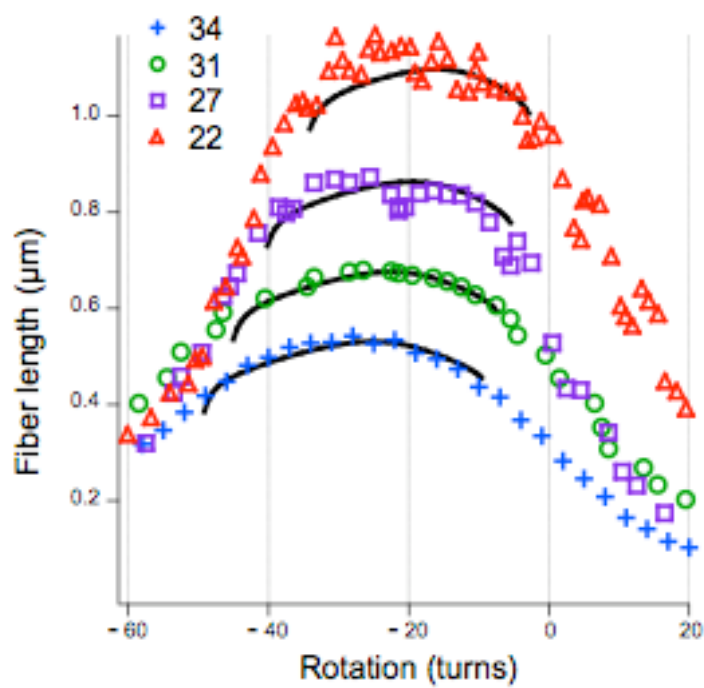

# Figure 7

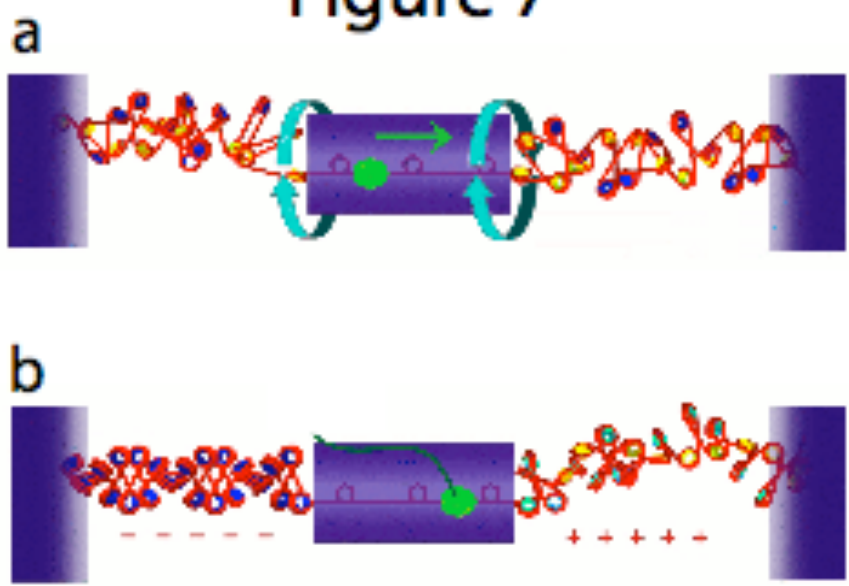